\documentclass[usenatbib,usegraphicx]{mn2e}

\newcommand{\aap}{\textit{A\&A}}

\newcommand{\apj}{\textit{ApJ}}

\newcommand{\apjl}{\textit{ApJ Lett.}}
\newcommand{\apjs}{\textit{ApJ Suppl.}}

\newcommand{\mnras}{\textit{MNRAS}}

\newcommand{\araa}{ARA\&A}

\newcommand{\cs}{c_{\rm s}}

\newcommand{\kms}{{\rm ~km~s}^{-1}}

\newcommand{\Mdense}{M_{\rm dense}}

\newcommand{\Mmax}{M_{\rm max}}
\newcommand{\mst}{m_{*}}
\newcommand{\Mstar}{{M_*}}
\newcommand{\Msun}{M_\odot}
\newcommand{\Mvir}{M_{\rm vir}}
\newcommand{\nlos}{n_{\rm LOS}}
\newcommand{\nsf}{n_{\rm SF}}

\newcommand{\pcc}{{\rm ~cm}^{-3}}

\newcommand{\Rs}{R_{\rm s}}
\newcommand{\soned}{\sigma_{\rm 1D}}
\newcommand{\ts}{t_{\rm s}}

\newcommand{\vrms}{v_{\rm rms}}

\newcommand{\VS}{V\'azquez-Semadeni}

\def\mathnew{\mathsurround=0pt}
\def\simov#1#2{\lower .5pt\vbox{\baselineskip0pt
    \lineskip-.5pt\ialign{$\mathnew#1\hfil##\hfil$\crcr#2\crcr\sim\crcr}}}
\def\simgreat{\mathrel{\mathpalette\simov >}}
\def\simless{\mathrel{\mathpalette\simov <}}

\title[Cloud Destruction by Stellar Feedback]{Molecular Cloud Evolution V. 
Cloud Destruction by Stellar Feedback} 

\author[Col\' \i n et al.]{Pedro Col\'in$^1$, \thanks{e-mail: p.colin@crya.unam.mx}
Enrique \VS\ $^1$, and Gilberto C.\ G\'omez$^1$\\
$^1$Centro de Radiostronom\'ia y Astrof\'isica, UNAM, 
Apartado Postal 72-3 (Xangari), 58089 Morelia, Mexico}
\begin{document}
\label{firstpage}
\maketitle

\begin{abstract}
We present a numerical study of the evolution of molecular clouds, from
their formation by converging flows in the warm ISM, to their
destruction by the ionizing feedback of the massive stars they form. We
improve with respect to our previous simulations by including a
different stellar-particle formation algorithm, which allows them to
have masses corresponding to single stars rather than to small clusters,
and with a mass distribution following a near-Salpeter stellar IMF.
We also employ a
simplified radiative-transfer algorithm that allows the stellar particles to feed
back on the medium at a rate that depends on their mass and the local
density. Our results are as follows: a) Contrary to
the results from our previous study, where all stellar particles
injected energy at a rate corresponding to a star of $\sim 10 ~\Msun$,
the dense gas is now completely evacuated from 10-pc regions around the
stars within 10--20 Myr, suggesting that this feat is accomplished
essentially by the most massive stars. b) At the scale of the whole
numerical simulations, the dense gas mass is reduced by up to an order
of magnitude, although star formation (SF) never shuts off completely,
indicating that the feedback terminates SF locally, but new SF
events continue to occur elesewhere in the clouds. c) The SF efficiency
(SFE) is maintained globally at the $\sim 10\%$ level, although
locally, the cloud with largest degree of focusing of its accretion flow
reaches SFE $\sim 30\%$. d) The virial parameter of the clouds approaches
unity {\it before} the stellar feedback begins to dominate the dynamics,
becoming much larger once feedback dominates, suggesting that clouds
become unbound as a consequence of the stellar feedback, rather than
unboundness being the cause of a low SFE. e) The erosion of the filaments 
that feed the star-forming clumps produces chains of isolated dense blobs 
reminiscent of those observed in the vicinity of the dark globule B68. 
\end{abstract}

\section{Introduction} \label{sec:intro}

Understanding how star formation (SF) proceeds in our Galaxy, and other
galaxies in general, is a key quest in astrophysics. In recent years, it
has become clear that the evolution and distribution of SF
in the Galaxy is intimately linked to the structure and evolution of the
molecular clouds (MCs) where it takes place \citep[see, e.g., the
reviews by][and references therein]{MK04, BP_etal07, MO07, VS13}. 

In a series of previous papers \citep[][]{VS_etal06, VS_etal07,
VS_etal10, VS_etal11}, we have investigated, alongside other groups
\citep{HP99, HP00, AH05, HA07, AH10, Hennebelle_etal08, Heitsch_etal05,
Heitsch_etal06, HH08, Heitsch_etal09, Banerjee_etal09}, the evolution of
MCs and of their SF activity, from their formation by condensation of
the atomic gas in the interstellar medium to their star-forming stages
and, in some studies, to their destruction by stellar feedback
\citep{VS_etal10}. Among the above studies, those including self-gravity
\citep{VS_etal07, VS_etal10, VS_etal11, HH08, Heitsch_etal09} have shown
that the coherent formation of large clouds (several tens of parsecs)
leads to the onset of global gravitational contraction throughout the
cloud, at a stage when the cloud is still mostly composed of atomic
hydrogen, with SF only starting several Myr later, when the cloud has
become mostly molecular.

This result, however, is in contradiction with the largely established
notion that MCs cannot be collapsing freely, since otherwise their
resulting SF rates (SFRs) would be up to two orders of magnitude larger
than observed \citep{ZP74}. A related property is that the observed SF
efficiency (SFE, the fraction of a cloud's mass that ends up in stars)
for whole giant molecular clouds (GMCs) is estimated to be of only a few
percent \citep[e.g.][]{Myers_etal86, Evans+09, FK13}. Hence, it is
generally believed that MCs must be in or near equilibrium, supported
against their self-gravity by supersonic turbulence, magnetic fields, or
some combination thereof.  Specifically, a number of SF theories have
appeared in recent years in which the underlying scenario is that MCs
are supported globally by turbulent pressure, while gravitational
collapses occur only locally, caused by the supersonic turbulent
compressions \citep[e.g.,] [see also the discussion by Federrath \&
Klessen 2012] {PN02, PN11, KM05, HC11}.

Of course, an alternative explanation has been known for over four
decades \citep[e.g.,][]{Field70, Whitworth79, Elmegreen83, Cox83,
Franco+94} for the low observed SFE of GMCs, namely that stellar
feedback, mainly from the ionizing radiation from massive stars, may
disrupt the clouds before they have converted much of their mass into
stars. In this scenario, there is no need to support the clouds against
their self-gravity. Also, this scenario becomes even more feasible in
view of the recent result, from numerical simulations of cloud formation
and evolution, that the collapsing clouds undergo {\it hierarchical
gravitational fragmentation} \citep{VS_etal09}. That is, the turbulent
density fluctuations, having larger mean densities than that of the
whole parent GMC, have shorter free-fall times and smaller Jeans masses,
and therefore the densest clumps begin to form stars earlier than the
rest of the cloud, and before the global collapse of the cloud terminates.

In \citet[][hereafter Paper I]{VS_etal10}, we carried out a first
attempt to numerically capture this phenomenology, by performing
simulations of cloud formation and evolution in the presence of
ionization heating from massive stars, which is considered to be the
main feedback mechanism affecting GMCs of masses up to $\sim 10^5
~\Msun$ \citep{Matzner02, KM09, Dale+12}. In Paper I,
the instantaneous, time-dependent SFE was defined as
\begin{equation}
\hbox{SFE}(t) = \frac{\Mstar(t)}{\Mdense(t) + \Mstar(t)},
\label{eq:SFE_def_old}
\end{equation}
where $\Mdense$ is the mass in dense gas ($n > 100 \pcc$) and $\Mstar$
is the total mass in stars. It was found in that paper that the
prescription for ionization-heating feedback used there was able to
maintain the SFE at the few-percent level throughout the evolution of
the cloud, while control simulations not including it reached SFEs
roughly an order of magnitude larger. Also, an analytical model
representing this scenario was recently presented by \citet{ZVC12}, and
shown to correctly describe several evolutionary properties of GMCs and
their SF activity.

However, one shortcoming of the feedback prescription used in Paper I
was that it assumed that the stars responsible for the feedback all
injected energy into the medium at a rate roughly corresponding to that
of a $\sim 10~\Msun$ star. This implied that the stellar feedback was
possibly overestimated for clouds forming low-mass stars, and
underestimated for clouds forming high-mass stars. In particular, Paper
I found that the GMC-like clouds could not be destroyed by the
feedback. Instead, \citet{Dale+12}, for example, have been able to
disrupt clouds up to $10^6~\Msun$ by means of ionization feedback. In
Paper I, the SFE was kept low because the conversion of dense gas into
stars was locally inhibited by the feedback, but not because the clouds
at large were destroyed. Only the local clumps were destroyed.

In the present paper, we improve on the numerical prescription used in
Paper I in two ways. First, we use a probabilistic SF prescription
instead of a fully deterministic one. As it turns out, this
probabilistic prescription allows us to produce a mass spectrum for
stellar particles, which can be tuned to resemble the Salpeter initial
mass function (IMF). Second, once armed with a
realistic stellar mass spectrum, we incorporate a mass-dependent
ionization heating prescription for the feedback from the stellar
particles produced in the simulations, applying a simplified description
of radiative transfer, neglected in Paper I. We expect that, with this
prescription, we can obtain a more realistic description of the effect
of ionization feedback on MCs of various masses.

The plan of the paper is as follows. In Sec.\ \ref{sec:model}, we
describe the numerical model, focusing in particular on the heating and
cooling functions employed (Sec.\ \ref{sec:heat_cool}), the
stellar-particle formation prescription (Sec.\ \ref{sec:SF_model}), the
feedback prescription (Sec.\ \ref{sec:Feedback}), and the refinement
criterion used (Sec.\ \ref{sec:refinement}). We next describe the
details and parameters of the simulations in Sec.\
\ref{sec:simulations}, and then present our results in Sec.\
\ref{sec:results}. In Sec.\ \ref{sec:discussion} we compare our results
with those of Paper I and discuss some of their implications. Finally,
in Sec.\ \ref{sec:conclusions} we present a summary and some conclusions.

\section{The Numerical Model} \label{sec:model}

The numerical simulations used in this work were performed using the
adaptive mesh refinement (AMR) + N-body Adaptive Refinement Tree code ART
\citep[]{KKK97,Kravtsov03}. In the following sections we describe the
adaptations we have performed to it for application to our problem of
interest.

\subsection{Refinement} \label{sec:refinement}

The numerical box is initially covered by a grid of $128^3$ (zeroth
level) cells. The mesh is subsequently refined as the matter
distribution evolves. The maximum allowed refinement level was set to
five, so that high-density regions have an effective resolution of
$4096^3$ cells, with a minimum cell size of 0.0625 pc. As in Paper I,
cells are refined when the gas mass within the cell is greater than
$0.32 \Msun$. That is, the cell size is refined by a factor of 2 when
the density increases by a factor of 8, so that, while refinement is
active, the grid cell size $\Delta x$ scales with density $n$ as $\Delta
x \propto n^{-1/3}$. Once the maximum refinement level is reached, no
further refinement is performed, and the cell's mass can reach much
larger values.

Note that this constant-cell-mass refinement criterion does not conform
to the so-called {\it Jeans criterion} \citep{Truelove+97} of resolving
the Jeans length with at least 4 grid cells. \citet{Truelove+97}
cautioned that failure to do this might result in spurious, numerical
fragmentation. However, we do not consider this a cause for concern
since, as will be described in Sec.\ \ref{sec:SF_model}, our star
formation prescription allows us to choose the stellar-particle mass
distribution, and tune it to a \citet{Salpeter55} value.

\subsection{Heating and cooling} \label{sec:heat_cool}

The main additional physical 
processes implemented in our simulations, and relevant to the physical
problem studied here are a) the cooling and heating of the gas; b) its
conversion into stars; c) the stellar feedback via ionization-like
heating, and d) the self-gravity from gas and stars.

We use heating ($\Gamma$) and cooling ($\Lambda$) functions of the form  
\begin{eqnarray} 
\Gamma &=& 2.0 \times 10^{-26} \hbox{ erg s}^{-1}\label{eq:heating}\\
\frac{\Lambda(T)}{\Gamma} &=& 10^7 \exp\left(\frac{-1.184 \times
10^5}{T+1000}\right) \nonumber \\
&+& 1.4 \times 10^{-2} \sqrt{T} \exp\left(\frac{-92}{T}\right)
~{\rm cm}^3.\label{eq:cooling}
\end{eqnarray}
These functions are fits to the various heating and
cooling  processes considered by
\citet{KI00}, as given by equation (4) of \citet{KI02}. As noted in
\citet{VS_etal07}, eq.\ (4) in \citet{KI02}
contains two typographical errors. The form used here incorporates the
necesary corrections, kindly provided by H.\ Koyama (2007, private
communication).  With these heating and cooling functions, the gas is
thermally unstable in the density range $1 \la n \la 10\ \pcc$.

\subsection{Star formation prescription: a probabilistic approach}
\label{sec:SF_model}  

In our simulations, SF is modeled as taking place in the densest
regions, defined by $n > \nsf$, where $n$ is the gas density, and $\nsf$
is a density threshold. If a grid cell meets this density criterion,
then a stellar particle (SP) of mass $m_*$ may be placed in the cell,
with probability $P$, every timestep of the coarsest grid. If the SP
{\it is} created, it acquires half of the mass of its parent cell, and
this mass is removed from the cell. Thereafter, the particle is treated
as non-collisional and follows N-body dynamics. No other criteria are
imposed. We set $\nsf = 9.2 \times 10^4
\pcc$, which corresponds to a cell mass of $0.78\ \Msun$ at the highest
refinement level. This fixes the minimum value for SP
masses at $0.39\ \Msun$. Note that, as in Paper I, our SPs differ from
the commonly-used sink particles \citep{Bate+95, Federrath+10}, mainly
in that our SPs are not allowed to accrete after they
form. Thus, we refrain from calling them as ``sinks'', and use the
nomenclature ``stellar particles'' instead. However, as we describe
below, our probabilistic approach to SP formation allows
us to obtain a realistic mass distribution (IMF) for them.


A few items are worth noting about our SF prescription. First, note
that, once the maximum refinement level is reached, no further
refinement is applied to a cell (cf.\ Sec.\ \ref{sec:refinement}) even
if its density keeps increasing. Moreover, since the creation of an SP
is a probabilistic event, the density of a cell where a gravitational
collapse is going on continues to increase until an SP forms in the cell.
Some authors have advocated the prescription that, once the maximum
refinement level has been reached, a sink particle is created at the
cell density that would correspond to the next refinement level
\citep[e.g.,] [] {Federrath+10} in order to always fulfill the Jeans
criterion and thus completely avoid spurious fragmentation
\citep{Truelove+97} until the sink particles are formed. However, we
forgo of this recommendation since, as will be seen in what follows, our
prescription allows us to {\it impose} the desired IMF of the SPs, and
thus artificial fragmentation is not a concern.

\begin{figure}
\includegraphics[width=0.45\textwidth]{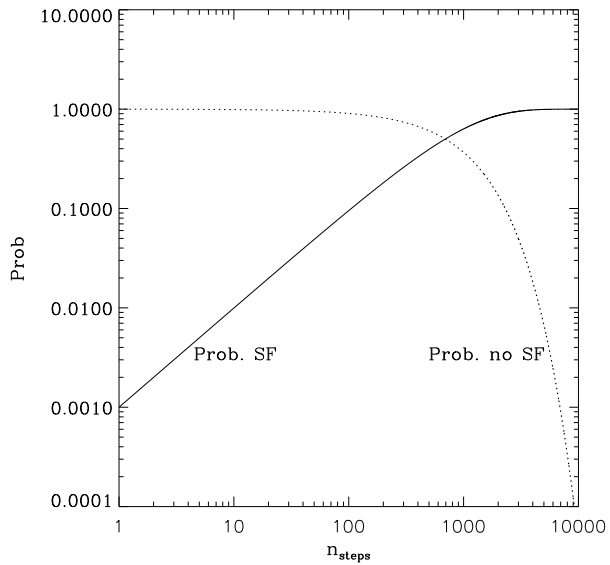}
\caption{Probability of having ({\it solid line}) and {\it not} having
formed ({\it dashed line}) a stellar particle (SP) in a cell that meets the
density criterion $n > \nsf$ after a certain number of coarse-grid 
time steps for $P=0.001$.
\label{fig:prob_sf}}
\end{figure}

\begin{figure}
\includegraphics[width=0.45\textwidth]{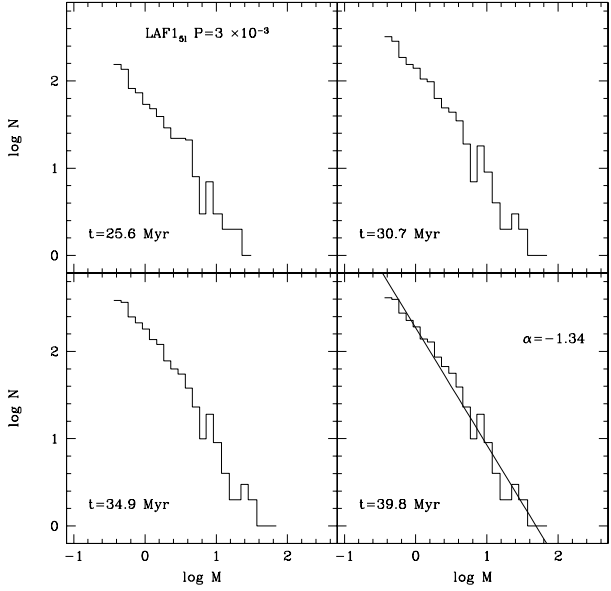}
\caption{Evolution of the spectrum of stellar masses for simulation
LAF1$_{5l}$. The spectrum at any epoch can be well fitted by a power law.
In the lower right panel, with a line, we also show this fit and the
value of its slope.
\label{fig:IMF}}
\end{figure}

The prescription we use implies that the longer it takes to form an SP
in a collapsing cell, the more massive the SP will be, because the
cell's density will be higher. The probability of {\it not} having
formed an SP after $n_{\rm steps}$ time steps is ${\cal P}_{\rm no} =
(1-P)^{n_{\rm steps}}$, while the probability of having formed it is
${\cal P}_{\rm yes} = 1 - (1-P)^{n_{\rm steps}}$. These probabilities
are shown in Fig.\ \ref{fig:prob_sf} for $P=0.001$.
Serendipitously, we have found that the resulting mass distribution of
the SPs is a power law, with an exponent that depends on
the value of $P$. Thus, $P$ is a control parameter that allows us to
generate a stellar mass spectrum with the desired slope. In Figure
\ref{fig:IMF} we show the evolution of the stellar mass spectrum in
simulation LAF1$_{5l}$ (cf.\ Sec.\ \ref{sec:simulations}) with
$P=0.003$. The slope of the function changes from $-1.21$ at $t=25.6$
Myr to $-1.34$ at the end of the evolution, thus hovering close to the
Salpeter value of $-1.35$. The most massive SP formed in
this simulation has $m_* =61\ \Msun$, while the least massive ones have
masses $m_* \sim 0.5~\Msun$. Note that we have no turnover of the IMF at
small masses, but this is inconsequential for our purposes, since we are
only interested in the feedback exerted by the stars on their parent
cloud, and the low-mass stars exert no significant feedback at the GMC
scale \citep[see, e.g., the review by][and references therein]{VS11}

We note that, because now the SPs form in cells whose density
is typically much larger than the threshold value $\nsf$, in the present
paper we choose $\nsf = 9.2 \times 10^4~\pcc$, to allow for a
sufficiently large number of SPs to form. This is significantly smaller
than the value used in Paper I, where SPs formed always at a density
very similar to $\nsf$. Moreover, we stress that, contrary to the
situation in our previous papers, our SPs now have masses corresponding
to individual stars rather than to small clusters, and so, in what
follows, we shall indistinctly refer to them simply as ``stars''.

An important concern is whether our probabilistic prescription
introduces a significant delay for the formation of massive stars, in
comparison to the relevant timescales in the simulations. To check for
this, we note that, according to a probabilistic sampling of the IMF, a
10-$\Msun$ star should appear after $\approx 106.4~\Msun$ of gas have
been converted to stars, so we can check whether the formation of such a
star is significantly delayed with respect to the time when this much
mass has been converted into stars in the simulations. We find that, in
run SAF1 (see Sec.\ \ref{sec:simulations}), SF starts at $t = 18.39$
Myr, while $106.4 \Msun$ worth of stars are reached at $t = 21.1$ Myr,
and a 10-$\Msun$ star appears at t = 21.3, so the time taken by the
simulation to form such a massive star coincides within less than 10\%
with the time needed for such a star to appear according to a
statistical sampling of the IMF. In run LAF1, these times are,
respectively, 18.74, 20.94, and 20.03 Myr, and so, in fact, a massive
star forms slightly earlier than the time at which $106.4 \Msun$ worth of
stars are present. So, we conclude that there is no significant delay
introduced by our prescription.

\subsection{Feedback prescription}
\label{sec:Feedback}  

Another important difference of our new feedback prescription, compared
to that in Paper I, is in the way we implement the ionization feedback
by massive stars. In Paper I, SPs injected thermal energy
only to the cell where they were located (hereafter, the ``stellar
cell''), at a rate high enough to produce a realistic HII
region\footnote{Because the thermal energy was dumped only in the cell
were the SP was formed, neighboring cells were heated by
numerical conduction.}. Instead, here we now model the birth and
evolution of HII regions by assigning a temperature of $10^4$ K to all
cells whose distance $d$ to the SP satisfies the condition
\begin{equation}
d < \Rs \equiv \left( \frac{3}{4 \pi} \frac{S_*}{\alpha \nlos^2} \right)^{1/3},
\label{eq:Rs}
\end{equation}
where $\Rs$ is the \citet{Stromgren39} radius, $S_*$ is the flux of
ionizing photons produced by the star, $\alpha = 3.0 \times 10^{-13}$
cm$^3$ s$^{-1}$ is the recombination coefficient, and $\nlos$ is a
characteristic particle number density along the line of sight between
the stellar cell and the test cell, which we discuss
below. If it turns out that $\Rs$ is smaller than the size of the
stellar cell, we simply set the temperature of this cell equal to $10^4$
K, and no further calculation is done. On the other hand, if $\Rs$ is
larger than the stellar cell's size, then it is necessary to determine
whether $d < \Rs$ or not. In principle, this poses a radiative transfer
problem since, as is well known, eq.\ (\ref{eq:Rs}) is valid only for
the case when the medium between the stellar and the test cells has a
uniform density. However, if the medium is not uniform, then the
photoionization-recombination balance must be computed along the line
joining the SP and the grid cell in question \citep[see,
e.g.,][]{Dale+07}, a procedure that can be quite computationally
expensive.

As a zeroth-order approximation to solve this problem, we opt for
choosing a value for $\nlos$ that can be deemed representative of the
typical density along the path from the SP to the grid
cell. Specifically, we take the geometric mean of the densities at these
two locations in the simulation, $\nlos = \sqrt{n_{\rm SP}
~n_{\rm test}}$. This approximation for $\Rs$ is, of course, crude, and
will miss, for example, shadowing effects due to intervening dense
clumps between the stellar and the test cells but, for the purpose of
modeling the large-scale dynamics of the MC containing these cells, we
consider it is sufficient. We jokingly refer to this scheme as a ``poor
man's radiative transfer'' (PMRT) scheme.

For the ionizing flux $S_*$, which depends on the SP's
mass, we use the tabulated data provided by \citet{Diaz_etal98}. Note
that only SPs with a mass greater than $1.9~\Msun$ inject
any significant ionizing feedback into the ISM, implying that only the
massive SPs influence the dynamics of the molecular
clouds.  Finally, note that we turn off the cooling for the cells whose
temperature is set to $10^4$ K. Otherwise, very dense cells would
radiate away their thermal energy very quickly. Their temperature is
held at $10^4$ K for a time $\ts$, which we assume
depends on the star's mass $\mst$ as
\begin{equation}
\ts = \left\{ \begin{array}{ll}
      \mbox{2 Myr} & \mbox{if $\mst \le 8 \Msun$};\\
      222 ~\mbox{Myr}~\left(\frac{\mst}{\Msun} \right)^{-0.95} & \mbox{if $\mst > 8 \Msun$}.\end{array} \right.
\label{eq:st_lifetime}
\end{equation}

For stars more massive than $8 \Msun$, this time is a fit to the stellar
lifetimes by \citet{Bressan+93}, while for stars with masses lower than that,
it represents the fact that the duration of the stellar-wind phase is
$\sim 2$ Myr, roughly independently of mass. This also means that we are
representing the effect of the winds and outflows of low mass stars by
an ionization prescription. While this is clearly only an approximation,
we do not expect it to have much impact on our calculations, since the
main source of feedback energy at the level of GMCs is the ionization
feedback from massive stars \citep{Matzner02}.

\begin{figure}
\includegraphics[width=0.45\textwidth]{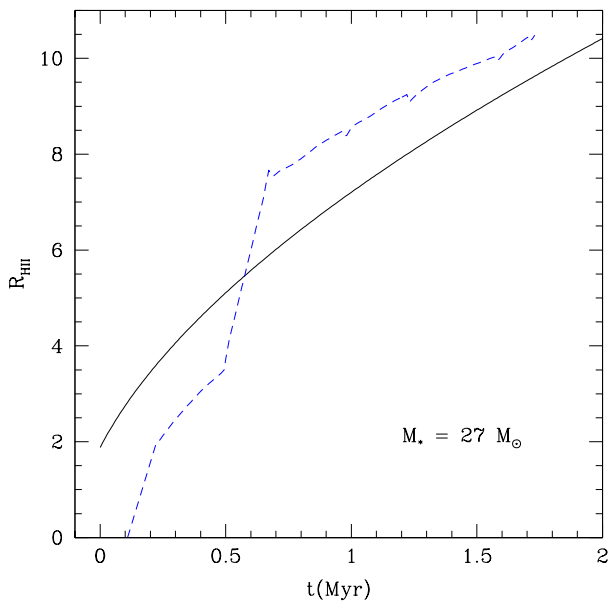}
\caption{Evolution of the HII region produced by a $27$-$\Msun$ star in a
box of 32 pc per side. The black solid line represents the analytic
solution given by eq.\ (\ref{eq:Rs_analytic}), while the blue dashed
line is the radius of the HII region in the simulation. Temperatures of
target cells located inside the Str\"omgren sphere, centered on the
SP, are set to $10^4$ K. The cooling is switched off in
these cells during the lifetime of the star.
\label{fig:Evolved_HII_Test}} 
\end{figure}

Our prescription for the ionization feedback for massive stars was
tested by running simulations of a box of 32 pc on a side, without
self-gravity, filled with gas at uniform temperature and density, of 42
K and $100 \pcc$, respectively. These simulations used a resolution of
$256^3$ cells, with the adaptive refinement switched off. A massive SP
($m_* = 27~\Msun$) was placed in the center of the box, and the system
was allowed to evolve freely.
Figure \ref{fig:Evolved_HII_Test} shows the expansion of the resulting
HII region, which is comprised of those cells with temperatures greater
than a few thousand degrees. Also shown in this figure is the well known
analytical solution \citep{Spitzer78},
\begin{equation}
R_s(t) = R_i \left( 1 + \frac{7}{4}\frac{c_s t}{R_i} \right)^{4/7},
\label{eq:Rs_analytic}
\end{equation}
where $R_i$ is the initial Str\"omgren radius and $c_s$ is the sound
speed. The numerical solution is seen to agree with the analytic one to
within $\sim 30$\%, an accuracy we consider sufficient, given our
interest only in the large-scale evolution of the clouds.

\section{The simulations} \label{sec:simulations}

Our simulations use the same initial 
setup as the runs in Paper I, which
represents the evolution of a region of 256 pc per side, initially
filled with warm gas at a uniform density of $n_0 = 1 \pcc$ and a
temperature $T_0 = 5000$ K, implying an adiabatic sound speed $\cs = 7.4
\kms$ (assuming a mean particle mass $\mu =1.27$). The full numerical
box thus contains $5.25 \times 10^5 \Msun$. In this medium, we make two
streams collide with a speed $v_{\rm inf} = 5.9 \kms$ each
(corresponding to a Mach number of 0.8 with respect to the unperturbed
medium) along the $x$-direction \citep[see Figure 1
of][]{VS_etal07}. The streams have a radius of 64 pc and a length of 112
pc each, so that the total mass in the two inflows is $9.0 \times 10^4
\Msun$. Note that the streams are completely contained within the box,
so that the compression they produce is a single event. There is no
continuous flow through the boundaries, as we use periodic boundary
conditions.
\begin{table*}
\caption{\sc Parameters of the Simulations}
\begin{tabular}{ccc}
\hline
Run     & $\vrms$       & Feedback \\
name	& [km s$^{-1}$] &	\\
\hline
SAF1 	& 0.1		& New prescription, full IMF	\\
LAF1 	& 1.7		& New prescription, full IMF	\\
LAF0	& 1.7           & Off         \\
LAFold  & 1.7           & Old prescription from Paper I  \\
LAF8 	& 1.7		& New prescription, max stellar mass $=8~\Msun$\\
LAF20 	& 1.7		& New prescription, max stellar mass $=20~\Msun$\\
\hline
\end{tabular}
\label{tab:run_params}
\end{table*}

On top of the inflow velocity we superpose a field of initial
low-amplitude turbulent velocity fluctuations, in order to trigger the
instabilities in the compressed layer that will cause it to fragment and
become turbulent \citep{Heitsch_etal05, VS_etal06}. As in Paper I, we
create this initial velocity fluctuation field with a new version of the
spectral code used in \citet{VPP95} and \citet{PVP95}, modified to run
in parallel in shared-memory architectures.  The simulations are evolved
for about 40 Myr.

The collision nonlinearly triggers a transition to the cold phase,
forming a turbulent, cold, dense cloud \citep{HP99, AH05, Heitsch_etal05,
Heitsch_etal06, VS_etal06}, consisting of a complex network of sheets,
filaments, and clumps of cold gas embedded in a warm diffuse substrate
\citep{AH05, HI06, HA07}. The complex as a whole quickly engages in
gravitational collapse \citep{VS_etal07}. Moreover, the local density
fluctuations become unstable and collapse in a shorter time than the
global time because they are embedded in a contracting medium and thus
have shorter collapse times \citep{Toala+13}. Eventually, they proceed
to forming stars, which then heat their environment, forming expanding
``HII regions'' that tend to disperse the clouds.

Although our results are based essentially on two simulations, a few
more runs were performed with a twofold purpose: to compare the old
prescription for feedback to the new one, and to assert the importance
of the most massive SPs in the disruption of the clouds. All simulations
but one were run using the ``large-amplitude'' (LA) initial velocity
fluctuations ($\vrms \sim 1.7\kms$) as opposed to the
``small-amplitude'' (SA) case ($\vrms \sim 0.1 \kms$) (see Paper
I). Clearly, more fragmentation and more complex cloud structures are
expected in the LA runs, thus causing them to produce somewhat smaller
clouds that resemble low- or intermediate-mass star-forming clouds. On
the other hand, the SA run allows us to consider a case of very high
coherence and uniformity, which tends to form a high-mass star-forming
region \citep{VS_etal09}, and furthermore resembles the conditions we
used in previous papers \citep{VS_etal07, VS_etal11}.

The nomenclature for the runs introduced in Table 1 continues to use the
acronyms LAF or SAF used in Paper I, where F stands for feedback. A
number 1 or 0 after the letter F means feedback is on or off,
respectively.  The test simulations are denoted with ``old'', 8 or 20
after the word LAF.

\section{Results} \label{sec:results}

\subsection{Evolution of the simulations} \label{sec:evolution}

\begin{figure*}
\includegraphics[width=0.95\textwidth]{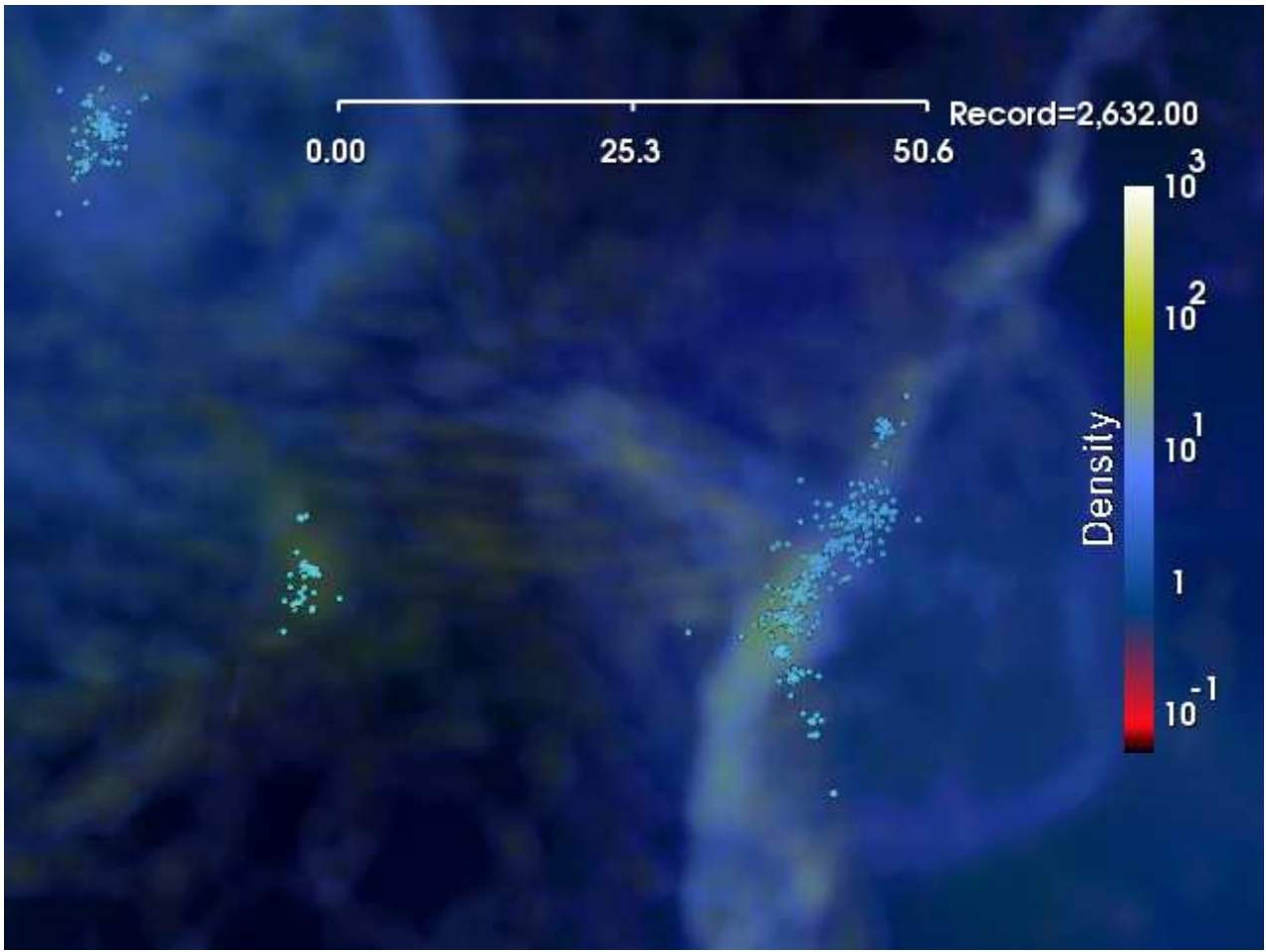}
\caption{View in projection of the central region of run LAF1 at $t
\approx 26.3$ Myr. The horizontal ruler shows a scale of 50.6 pc and we
have indicated with bluish dots the stellar particles. Three
star-forming clouds are seen, one at the upper left corner (Cloud 1),
one to the right of the image center (Cloud 2), and uncharted,
low-mass cloud at the left of the center. Shells expanding away from
Clouds 1 and 2 can be seen at this time.}
\label{fig:star_forming}
\end{figure*}

 The simulations performed here behave very similarly to previous
simulations with similar setups, except for the ultimate fate of the
individual clouds. In
particular, our SAF1 run is very similar to run L256$\Delta v$0.17 in
\citet{VS_etal07}, the run presented by \citet{Banerjee_etal09},
and the SAF0 and SAF1 runs in Paper I. The main feature of these runs is
that, because the initial velocity fluctuations are very mild, the flow
collision creates a large, coherent pancake-like structure of cold,
dense gas, which soon begins to undergo global gravitational
collapse. However, a recent study, in which the parameters of the flow
collision are varied \citep{Rosas_etal10}, has shown that the coherence
of the collapse may be lost in the presence of stronger initial
fluctuations.  In such cases, smaller clouds appeared to be less
strongly gravitationally bound, with the effect of decreasing the
SFE. This behavior was also observed in the LA simulations of Paper I,
and also occurs in the LAF1 run of the present paper. In this case, the
cloud formed by the initial flow is much more irregular in shape, and
much more fragmented and scattered over the simulation volume. As a
result, SF also occurs in a much more scattered manner, and the SFEs are
in general smaller in the LA runs than in their SA counterparts. Figure
\ref{fig:star_forming} shows an image, in projection, of the gas density
field of a region of run LAF1 that encompases the clouds we discuss in
the remainder of the paper: Cloud 1 (upper left corner), Cloud 2
(right off-center), and an uncharted third cloud (left off-center) in
run LAF1 at $t \approx 26.3$ Myr. In the figure, stellar particles are
shown as bluish dots.

However, in general a common pattern is followed by all simulations: the
transonic converging flows in the diffuse gas induce a phase transition
to the cold phase of the atomic gas. The newly formed dense gas is
highly prone to gravitational instability. This can be seen as
follows. The thermal pressure at our initial conditions is 5000
K$\pcc$. From Fig.\ 2 of \citet{VS_etal07}, it can be seen that the thermal
balance conditions of the cold medium at that pressure are $n \sim 130
\pcc$, $T \sim 40$ K. At these values, the Jeans length and mass are
$\sim 7$ pc and $\sim 640
\Msun$, respectively. These sizes and masses are easily achievable by a
large fraction of the cold gas structures, which can then proceed to
gravitational collapse and form stars. Moreover, the ensemble of these
clumps may also be gravitationally unstable as a whole, the likelihood
of this being larger for greater coherence of the large-scale pattern.

Regions of active star formation form in SAF1 and LAFs runs by the
{\it gravitational} merging of pre-existing smaller-scale clumps, which,
altogether, form a larger-scale GMC.

\begin{figure}
\includegraphics[width=0.45\textwidth]{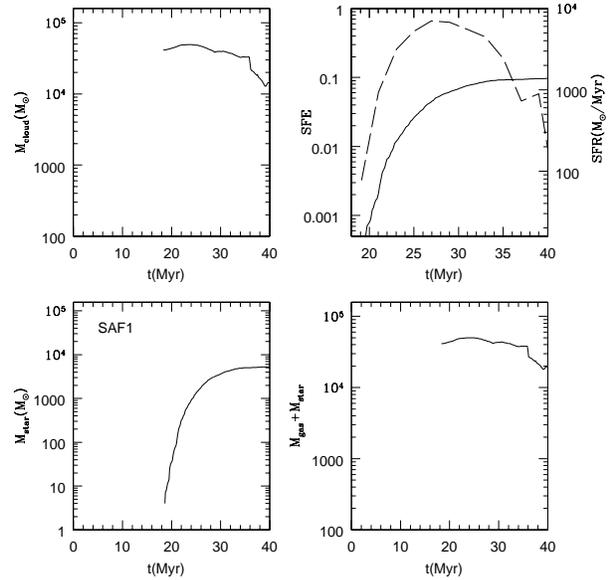}
\caption{Evolution of the mass in dense gas (top left
panel), the star formation rate (SFR; dashed line) and efficiency (SFE;
solid line) (top right panel), the mass in SPs (bottom left panel), and
the mass in stars plus dense gas (bottom right panel) in the whole
box in run SAF1. As a result of the destruction of the dense gas the
mass in SPs as well as the SFE reaches a maximum before the end of the
evolution.
\label{fig:Evolution_SAF1_Box}} 
\end{figure}

\subsection{Cloud Evolution in runs LAF1 and SAF1} \label{sec:LAF1_SAF1}

\begin{figure}
\includegraphics[width=0.45\textwidth]{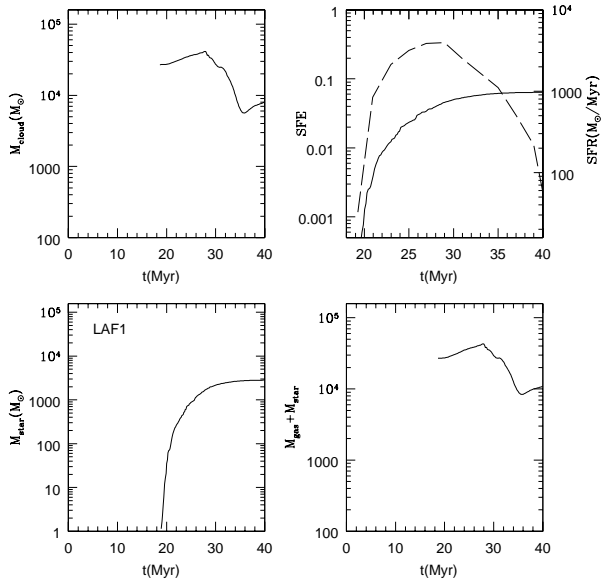}
\caption{Evolution of the mass in dense gas (top left
panel), the SFR (dashed line) and SFE (solid line; top right panel), the
mass in SPs (bottom left panel), and the mass in stars plus dense
gas (bottom right panel) in the whole box for run LAF1. Unlike the SA
run, here the star formation does not stop abruptly, and instead a slow
decline in the star formation rate is observed.
\label{fig:Evolution_LAF1_Single_Box}} 
\end{figure}

The new schemes for the probabilistic star formation recipe and for the
ionization feedback by massive stars were used to run the SAF1 and LAF1
simulations. Figures \ref{fig:Evolution_SAF1_Box} and
\ref{fig:Evolution_LAF1_Single_Box} show, for the full simulated box, the
evolution of the mass in dense gas ($n > 100\ \pcc$), the mass in stars,
the mass in both components, and the star formation rate (SFR) and
efficiency (SFE), in the SAF1 and LAF1 runs, respectively. In contrast
to Paper I, we now define the instantaneous SFE as
\begin{equation}
\hbox{SFE}(t) = \frac{\Mstar(t)}{\Mmax(t) + \Mstar(t)},
\label{eq:SFE_def}
\end{equation}
where $\Mmax$ is the maximum mass in dense gas reached between the start
of the simulation and time $t$. This is because in the present
simulations the clouds are eventually dispersed, and thus the SFE
evaluated with the instantaneous dense gas mass approaches unity at late
stages of evolution, but not because all the dense gas has been
converted to stars, but rather because the remaining gas is evaporated
by the stars. Our definition gives us instead an approximate measure of
the net SFE, that is, the fraction of the dense gas mass ever present in
a given volume and over a certain time interval that is converted into
stars.

In both runs, the evolution of the mass in dense gas is similar (see the
top left panels of Figs.\ \ref{fig:Evolution_SAF1_Box} and
\ref{fig:Evolution_LAF1_Single_Box}): first, dense gas starts to
accumulate as the evolution proceeds until it reaches a maximum, then
the effect of the radiation feedback is such that it overcomes the
buildup of dense gas by gravitacional accretion. Also, in both runs,
SPs continue to form after the maximum in the mass in
dense gas is reached, albeit at a significantly declining rate. Later
in the evolution, in the LAF1 run, the mass in dense gas begins to
increase again.  This time the density does not reach the threshold for
star formation and thus no new SPs are formed (see bottom
left panels of Figs.\ \ref{fig:Evolution_SAF1_Box} and
\ref{fig:Evolution_LAF1_Single_Box}).
\begin{figure}
\includegraphics[width=0.45\textwidth]{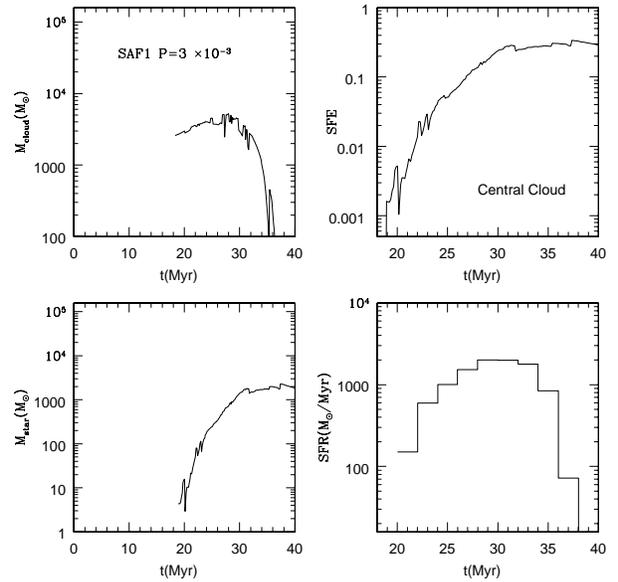}
\caption{Evolution of the mass in dense gas (top left
panel), the SFE (top right panel), the mass in SPs (bottom left panel),
and the SFR (bottom right panel) in the Central Cloud in run
SAF1. Ioinization feedback is so efficient that the cloud only lives
about 10 Myr. Interestingly, the stellar cluster is almost dispersed in
the next 10 Myr of evolution.
\label{fig:Evolution_Central_Cloud}} 
\end{figure}

In run SAF1, the largest star-forming region forms close to the center
of the box, due to the coherent collapse of the entire sheet-like cloud
formed by the collision. As in Paper I, we refer to this region as ``the
Central Cloud''. We enclose the cloud in a cylinder of radius 10 pc and
length 20 pc with its center located at the instantaneous minimum
of the potential within the cylindrical region, implying that the
cylinder moves in time following the cloud. Figure
\ref{fig:Evolution_Central_Cloud} shows the evolution of the 
mass in dense gas, the SFE, the mass in stars, and the SFR in this
cylinder. This figure is similar to Fig.\
\ref{fig:Evolution_SAF1_Box} (or Fig.\
\ref{fig:Evolution_LAF1_Single_Box}) except that the bottom right panel
now shows the evolution of the SFR in the cylinder instead of the
evolution of the gas-plus-stars mass. Unlike what happens with the whole
box, where some dense gas still remains by the end of the evolution,
here we witness the complete dispersal of the cloud from this region. In
addition, in the lower left panel we see that the mass in SPs also
decreases by the end of the evolution. Because our SPs have no winds and
do not explode as supernovae, this can can only mean that the stellar
cluster, formed from the dense gas mass of the cloud, is being dispersed
as well (that is, its constituent stars are leaving the cylinder that
initially contained the cloud).
\begin{figure}
\includegraphics[width=0.45\textwidth]{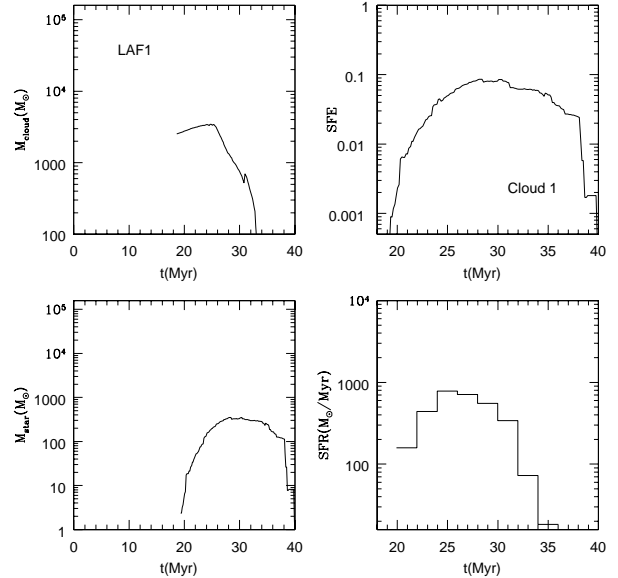}
\caption{Evolution of the mass in dense gas (top left
panel), the SFE (top right panel), the mass in SPs (bottom left panel),
and the SFR (bottom right panel) in Cloud 1 of run LAF1.
\label{fig:Evolution_LAF1_Single_Cloud1}}
\end{figure}

\begin{figure}
\includegraphics[width=0.45\textwidth]{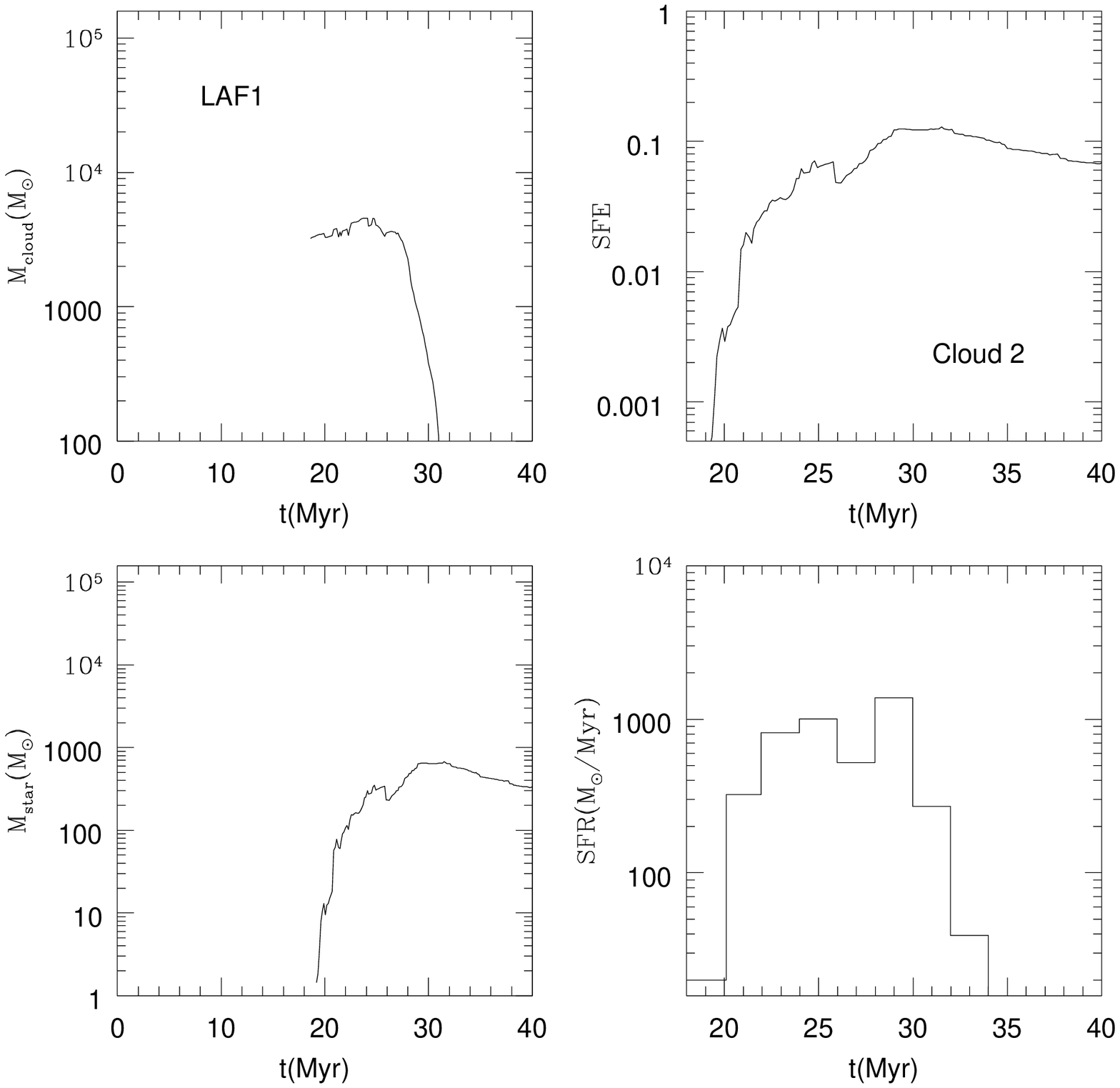}
\caption{Evolution of the mass in dense gas (top left
panel), the SFE (top right panel), the mass in SPs (bottom left panel),
and the SFR (bottom right panel) in Cloud 2 of run LAF1.
\label{fig:Evolution_LAF1_Single_Cloud2}}
\end{figure}

\begin{figure*}
\includegraphics[scale=0.35]{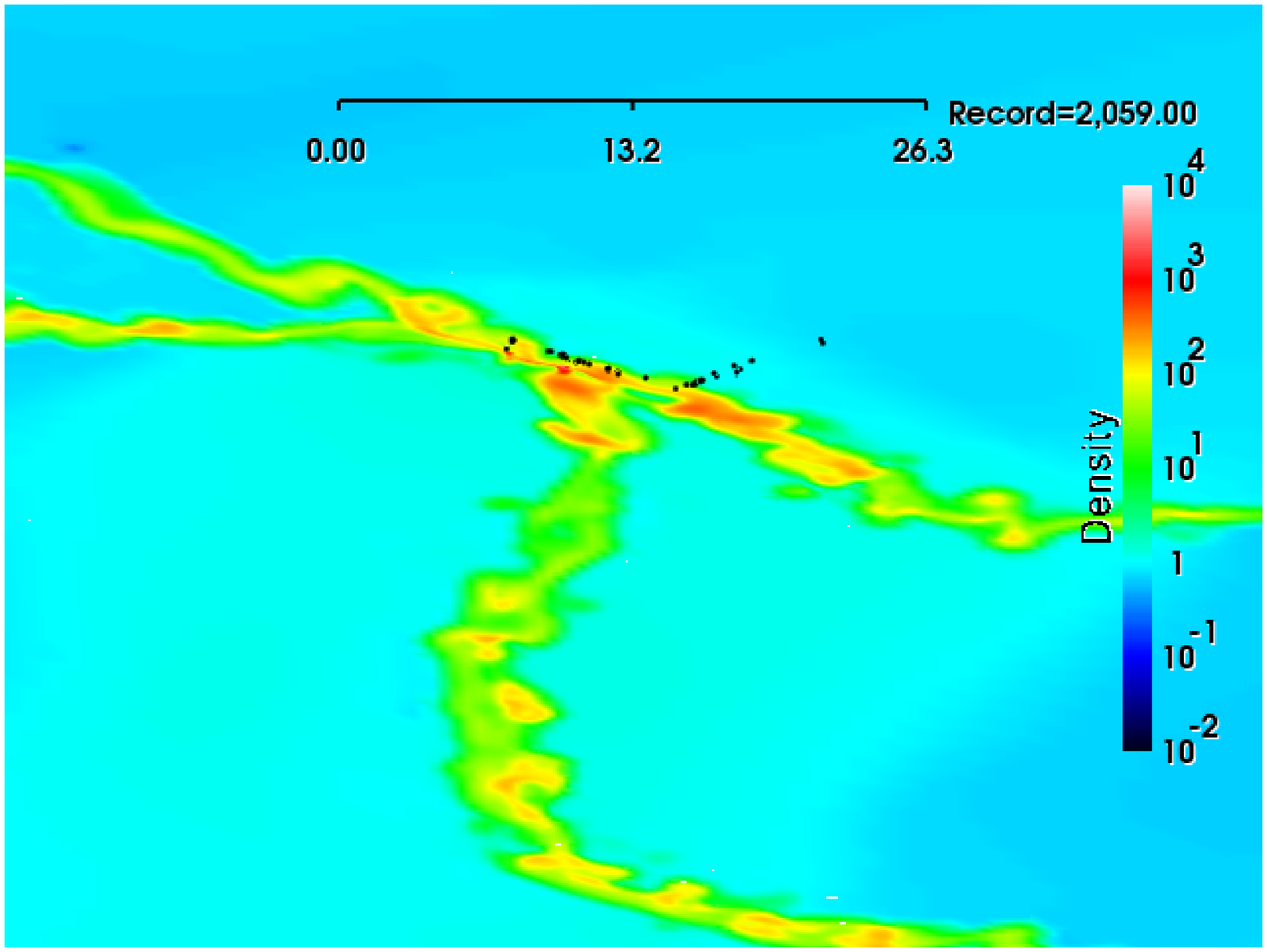}
\includegraphics[scale=0.35]{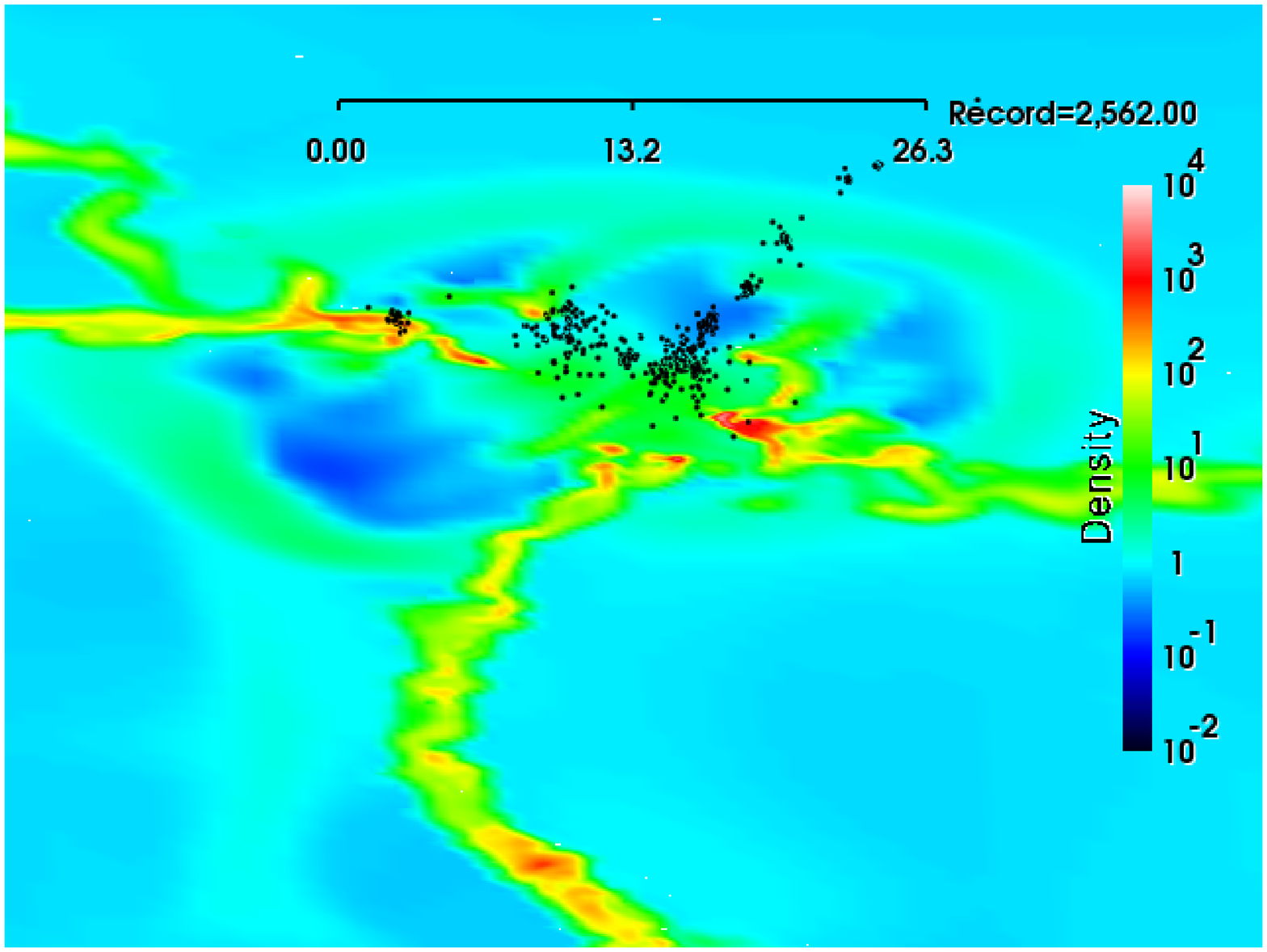}
\includegraphics[scale=0.35]{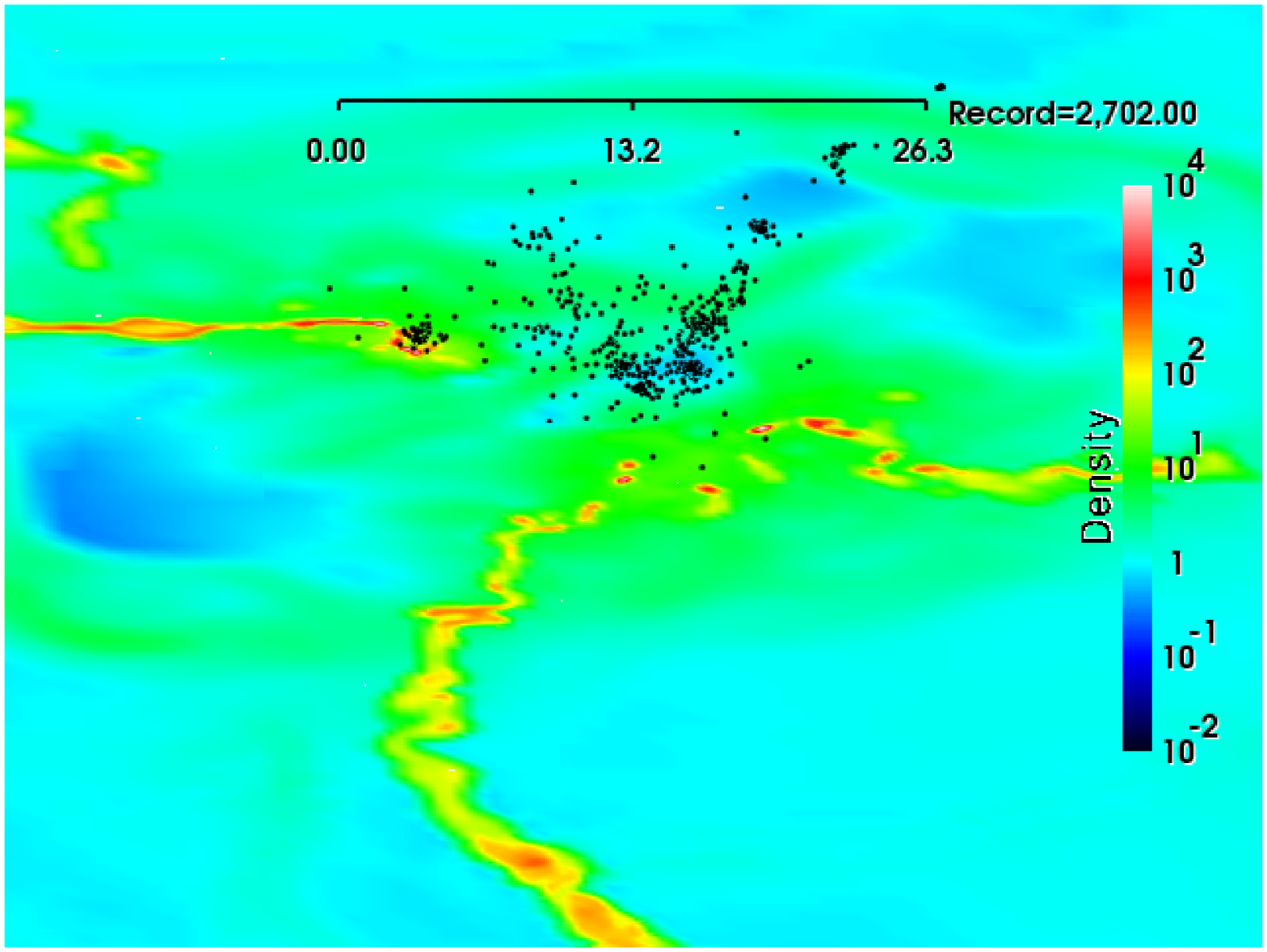}
\includegraphics[scale=0.35]{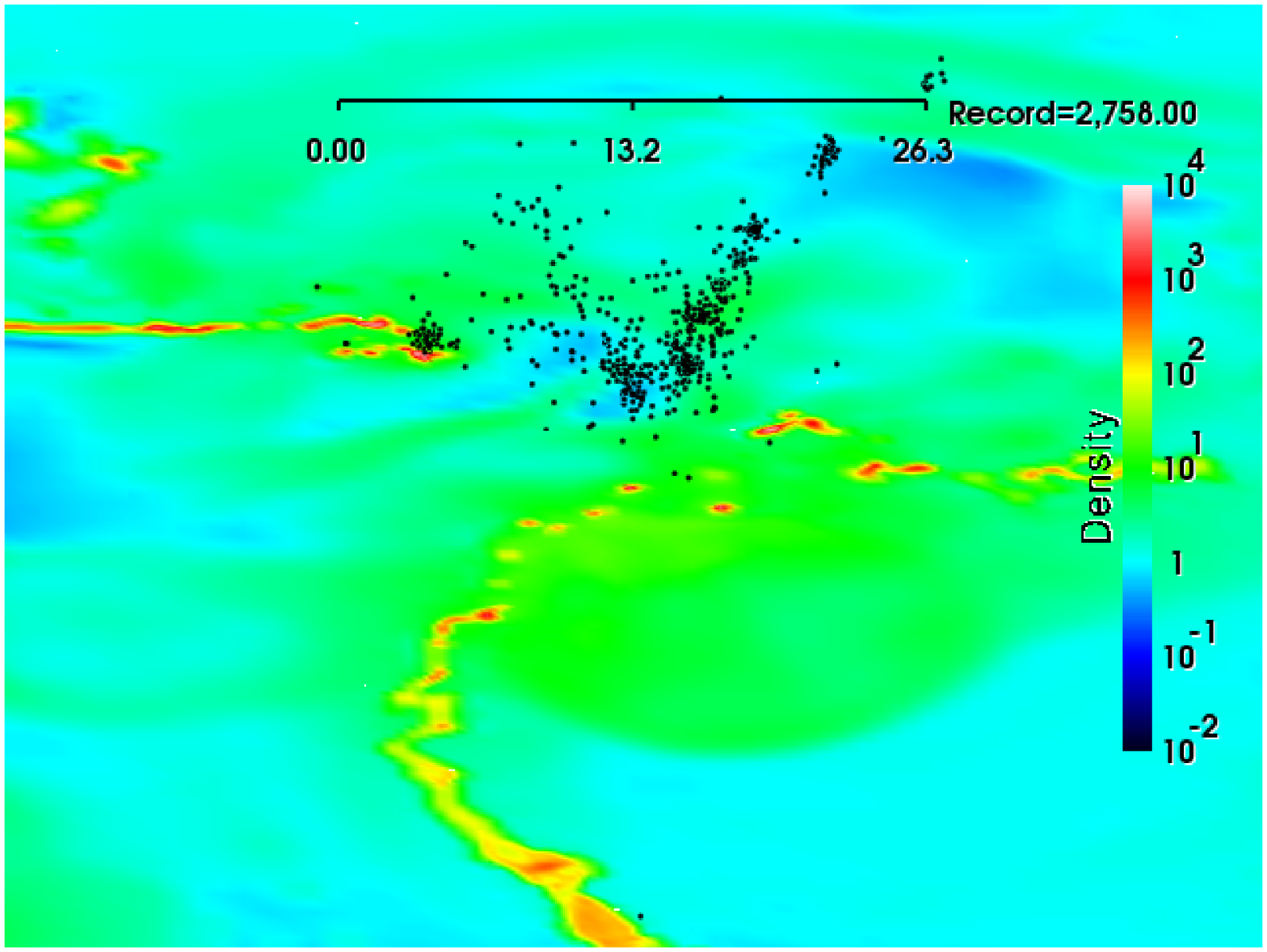}
\includegraphics[scale=0.35]{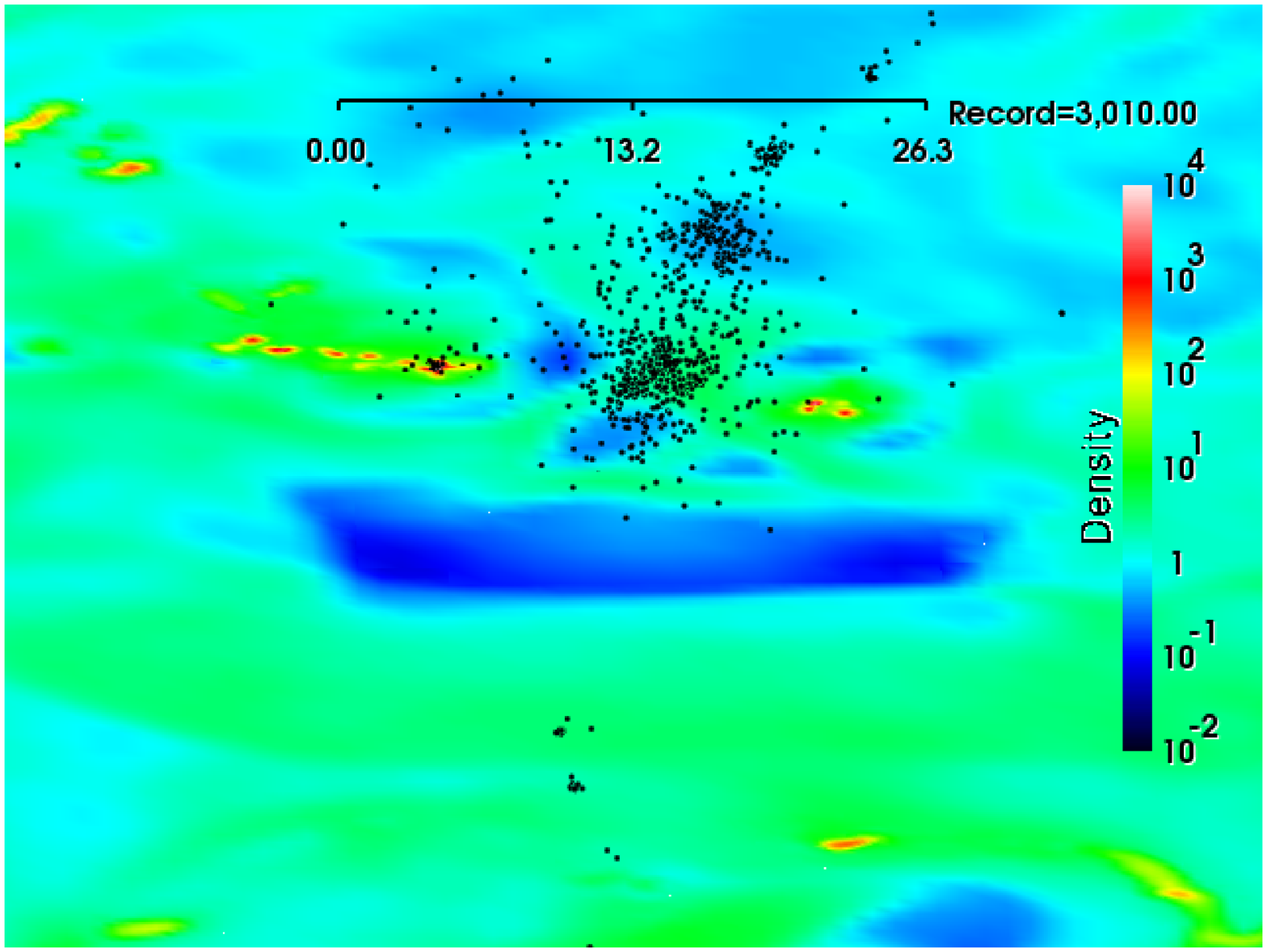}
\includegraphics[scale=0.35]{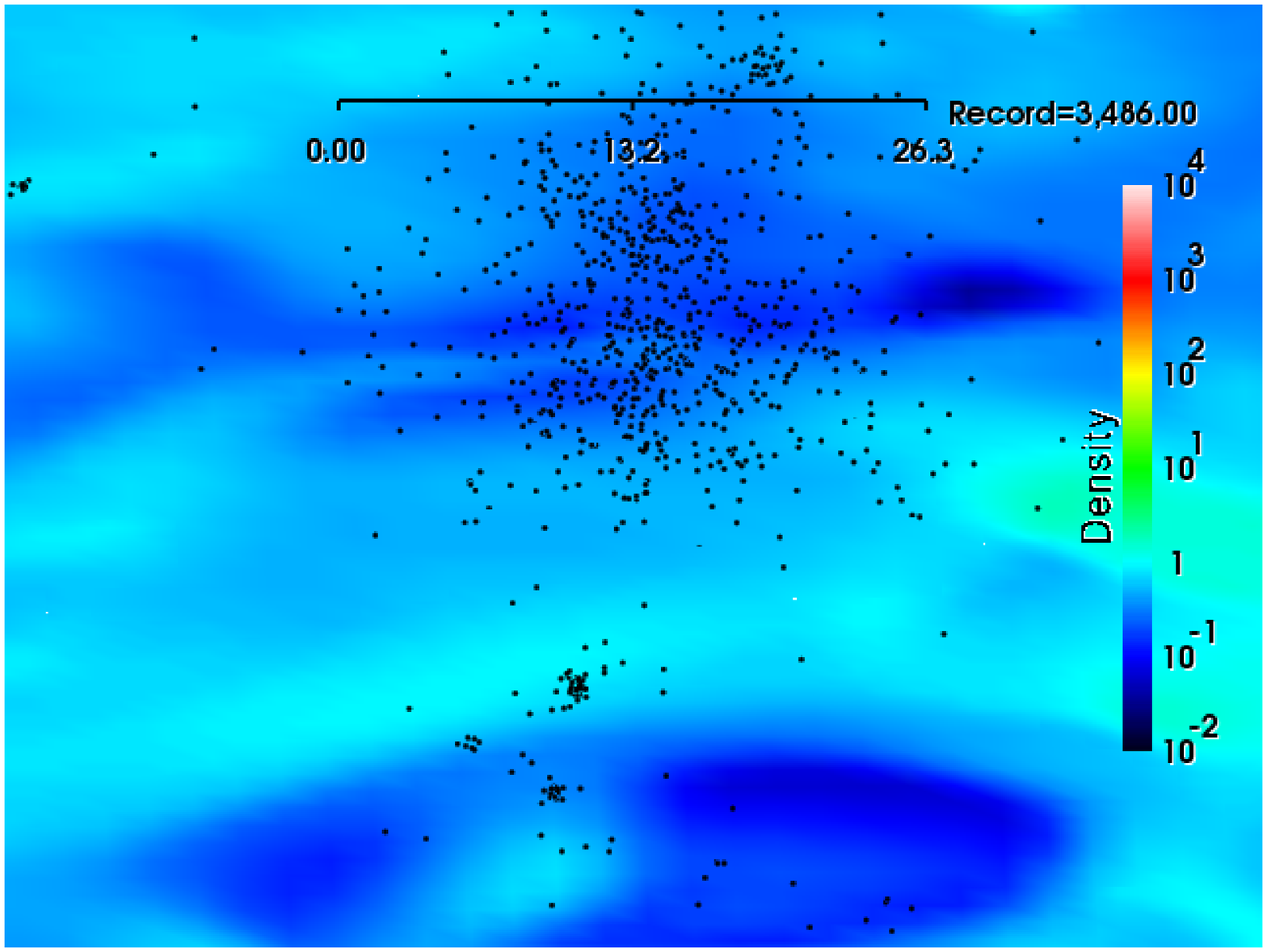}
\caption{Cross-section images of the density field in the neighborhood
of Cloud 2 in run LAF1, at times (in Myr) 20.59 (top left), 25.62 (top right),
27.02 (middle left), 27.58 (middle right),
30.1 (bottom left), and 34.86 (bottom right), showing the dispersal of
the cloud. The black dots show the stellar particles (SPs). The
horizontal ruler shows a scale of 26.3 pc. Note that the density field
is shown on an inclined cross section through the simulation, but the
SPs are shown in 3D space, so all particles in front to the density plane
can be seen. Note the complete dispersal of the cloud within 15 Myr.
\label{fig:images_cloud2}} 
\end{figure*}

In the case of run LAF1, the clouds identified as Cloud 1 and Cloud 2 in
Paper I are also used here to study the evolution of the cloud's
mass and its star formation activity. In paper I, we identified the
centers of the clouds visually\footnote{Coordenates of the Cloud 1 and
Cloud 2 are (x,y,z)= (100.0,140.0,150.0) and (150.0,115.0,105.0),
respectively, in pc.} and enclosed them in a cylinder of the same
dimensions as that used with the Central Cloud. Here, we use cylindrical
regions of the same size to locate the minima of the potential and place
the centers of the cylinders there at each time. As in Fig.\
\ref{fig:Evolution_Central_Cloud}, Figs.\
\ref{fig:Evolution_LAF1_Single_Cloud1} and
\ref{fig:Evolution_LAF1_Single_Cloud2} also show the evolution of the
mass in dense gas, the SFE, the mass in stars, and the SFR for Cloud 1
and Cloud 2, respectively.  As with the Central Cloud in run SAF1, Cloud
1 is also destroyed, and no stars are left inside the cylinder where the
cloud initially was; that is, the stellar cluster associated with the
cloud is dispersed, in $\sim 12$ Myr. Cloud 2 is also completely
destroyed, but unlike Cloud 1, here we still can find SPs inside its
corresponding cylinder, as it was the case for the Central Cloud. Figure
\ref{fig:images_cloud2} shows the evolution in the neighborhood of Cloud
2 over 15 Myr, illustrating these results.

A final remark is that Clouds 1 and 2 evolve essentially
independent from one another during the first several Myr after the
onset of SF. As seen in Fig.\ \ref{fig:star_forming}, they are separated
by nearly 60 pc. So, the shells expanding from them, at a speed of
roughly $10 \kms$, will reach the other region only after some 6
Myr. Moreover, as will be discussed in Sec.\ \ref{sec:most_massive},
only stars more massive than $20 \Msun$ are really effective in
destroying the clouds, and these only form several Myr after the onset
of SF, when a large enough mass has been converted to stars that such
massive stars are expected to form from a random sampling of the
IMF. For example, in run LAF1, a 30-$\Msun$ star forms at $t = 23.5$
Myr; that is, $\sim 5$ Myr after the onset of SF. Thus, the effect of one
star-forming region on the other is only expected to be important after
$\sim 10$ Myr, at which time the local effect of these massive stars
will have had plenty of time to act. Thus, we conclude that the effect
of one cloud on the other is negligible compared to that of the local
SF. However, the effect of one region on the other may be important when
only one of the two clouds manages to form massive stars.

\subsection{Evolution of the virial parameter} \label{sec:alpha}

One important parameter of molecular clouds is the so called
turbulent $\alpha$ parameter, defined as
\begin{equation}
\alpha \equiv 2 K/|W|,
\label{eq:alpha_def}
\end{equation}
where $K = 3\soned^2 M/2$ is the (turbulent) kinetic energy, with
$\soned$ the one-dimensional velocity turbulent dispersion and $M$ the
cloud's mass, and $W$ is the gravitational energy of the cloud,
neglecting environmental contributions \citep[see, e.g.,][]{BP06}. For a
spherical cloud of uniform density, $W = -3 GM^2/5R$, and eq.\
(\ref{eq:alpha_def}) becomes
\begin{equation}
\alpha = \frac{5 \soned^2 R}{GM} \equiv \frac{\Mvir}{M},
\label{eq:alpha_sphere}
\end{equation}
where the identity defines the {\it virial mass} as $\Mvir \equiv 5
\soned^2 R/G$. 

For a cloud in virial equilibrium, $\alpha =1$, and the fact that clouds
are often found to have values of the virial parameter near unity
\citep[or masses close to the virial mass; e.g.,][]{Heyer+09} is
generally interpreted as a signature of the clouds being in near virial
equilibrium, while it is often stated that clouds strongly dominanted by
self-gravity should have $\alpha \ll 1$. However, this would be true
only if the turbulent motions could be clearly separated from the
infalling motions that must develop in a collapsing cloud, a feat that
is very difficult to accomplish in practice. Moreover, it has been
recently pointed out by \citet{BP+11} that the free-fall velocities are
of the same order of magnitude as the virial turbulent motions since,
after all, both types imply kinetic energies comparable to the
gravitational energy. 

\begin{figure}
\includegraphics[width=0.45\textwidth]{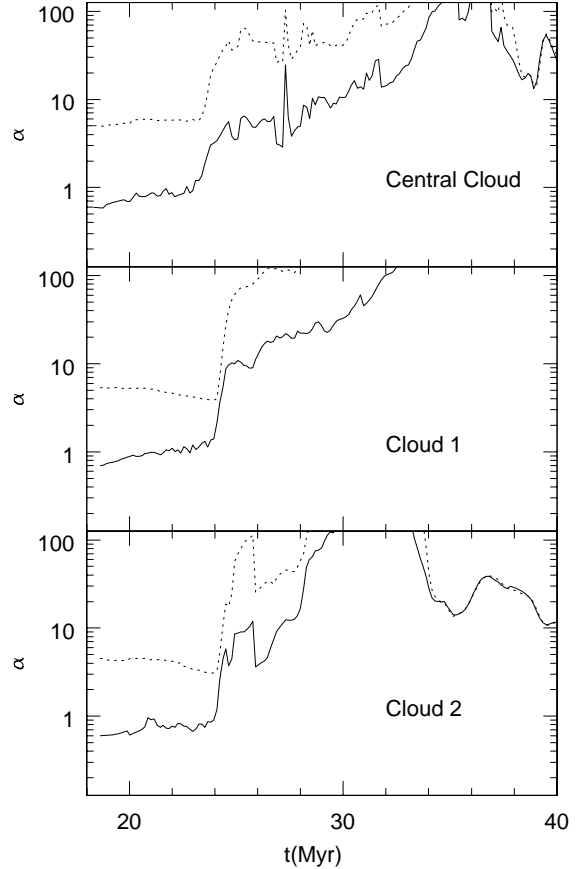}
\caption{Evolution of the alpha parameter for the three clouds
for runs SAF1 and LAF1. {\it Solid} lines
correspond to the density-weighted velocity dispersion, while {\it
dotted} lines correspond to the volume-weighted one.}
\label{fig:Alpha_parameter} 
\end{figure}

In Fig.\ \ref{fig:Alpha_parameter} we show the evolution of the $\alpha$
parameter for our three sample clouds from the onset of SF till the end
of the simulations, using either the density-weighted velocity
dispersion (which highlights the dense gas; solid lines) or the
volume-weighted velocity dispersion (which tends to highlight the
diffuse warm gas, since it occupies a larger fraction of the volume;
dotted lines). Interestingly, we see that, for all three clouds,
$\alpha$ for the dense gas is very close to unity, and in fact,
continues to approach it until the time when sufficiently massive stars
begin destroying the clouds, at which point it becomes much larger than
unity. Conversely, for the diffuse, warm gas, $\alpha$ is significantly
larger than unity at all times, although it becomes even larger when the
massive stars begin to drive the motions.

\subsection{Feedback scheme comparison}

In the feedback scheme of Paper I, HII regions were created by the
injection of thermal energy from SPs.  The energy was deposited entirely
in the cell where the stars were located, and thus neighboring cells
were heated exclusively by conduction, rather than by radiative
heating. The value of the rate at which the energy was dumped was chosen
so as to produce reasonably realistic HII regions. Additionally, the
cooling in the heated cell was turned off, since otherwise most of
energy would be radiated away in these initially very dense cells. Thus,
it is not feasible to directly compare the results of our new
simulations, with the PMRT scheme used in the present paper. However, it
is important to compare the old prescription with the new one used in
the present paper in a controlled manner, to assess the differences
induced by the prescription, in addition to the differences induced by
the presence of a stellar IMF. Therefore, we have run another LAF-type
simulation, labeled LAFold, which uses the old feedback
prescription from Paper I, and in which all SPs with $M_* > 10\ \Msun$
inject thermal energy at a rate equal to that used in Paper I. 

Figure \ref{fig:Evolution_LAF1_Box} shows the evolution of the mass in
dense gas, the SFE, the stellar mass, and the mass in dense gas plus
stars for runs LAF1 (black solid lines), LAF0 (red dotted lines)
and LAFold (blue, short-dashed lines), together with two other runs to be
discussed in the next section (see Table \ref{tab:run_params}). We see
that the LAFold run does not disrupt the clouds, in line with the results
of Paper I. 

\subsection{The role of the most massive stars in the destruction
of the clouds} \label{sec:most_massive}

\begin{figure}
\includegraphics[width=0.45\textwidth]{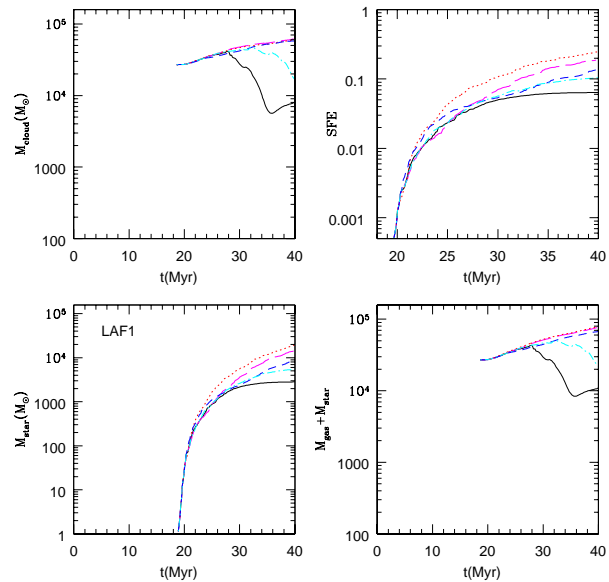}
\caption{Evolution of the mass in dense gas (top left
panel), the SFE (top right panel), the mass in SPs
(bottom left panel), and the mass in stars plus dense gas (bottom
right panel) in the whole numerical box of each of the five LAF runs in
the whole box: run LAF1 (black solid lines); run LAF0, with no feedback
(red dotted lines); run LAFold (blue short-dashed lines), with the
feedback prescription from Paper I; run LAF8 (magenta long-dashed
lines), with the new feedback prescription but with all SPs with $M_* >
10\ \Msun$ ionizing their surroundings as if they had a mass of $8
~\Msun$; and finally, run LAF20 (cyan dot-dashed lines), in which all
SPs with $M_* < 20\ \Msun$ feed back according to their masses, but SPs
with $M_* \ge 20\ \Msun$ feed back as if they were a star of $20\
\Msun$.
\label{fig:Evolution_LAF1_Box}}
\end{figure}

To assert the importance of the feedback of the massive stars 
in the destruction of the clouds, two extra LAF models were run,
labeled LAF8 and LAF20. LAF8 (magenta long-dashed lines in Fig.
\ref{fig:Evolution_LAF1_Box}) is a run 
with the new feedback prescription, but with all SPs
with $M_* > 10\ \Msun$ ionizing their surroundings {\it as if they
were a star of $8\ \Msun$}. LAF20 (cyan dot-dashed lines), moreover,
is a run similar to LAF1 but in this case the feedback ``saturates''
at $20\ \Msun$; that is, all SPs with $M_* < 20\ \Msun$
have the feedback they should according to their mass, but those
SPs with $M_* \ge 20\ \Msun$ exert a feedback as if they were
a star of $20\ \Msun$. 

From Fig.\ \ref{fig:Evolution_LAF1_Box} we see that, like run LAFold,
run LAF8 does not destroy the clouds; the mass in dense gas in the whole
simulated box (and in the individual clouds, not shown) continues to
increase. Run LAF20 is an intermediate case between those runs in which
clouds are not destroyed and run LAF1: in run LAF20, the mass in dense
gas reaches a peak before the end of the evolution. Because the feedback
in run LAF20 is not as strong as it is in run LAF1, this maximun is
reached few Myr later.  This experiment demostrates that stars with $M_*
\ge 20\ \Msun$ are crucial for the destruction of clouds of masses up to
a few times $10^4 ~\Msun$.

\subsection{Formation of ``dark globule'' chains} \label{sec:globules}

An interesting feature of run LAF1 is that, while the massive stars are
in the process of evaporating the dense gas, the filaments that feed the
cluster-forming clump are eroded, being destroyed first where the
densities are lowest. These filaments contain dense clumps, which are
more resilient to the effect of the ionising heating than the rest of
the filaments, and thus they remain for some time after the filamentary
structures have been destroyed. This process thus leaves behind chains
of dense blobs, strongly reminiscent of those observed, for example, in
the vicinity of the famous dark globule B68 \citep[e.g.,][] {Roman+10},
as can be seen in the top-right and middle (left and right) panels of
Fig.\ \ref{fig:images_cloud2}. In a forthcoming paper we plan to examine
in detail the similarities between the surviving blobs in our
simulations and the dark globules in the vicinity of {\sc Hii} regions.

\section{Discussion} \label{sec:discussion}

\subsection{Comparison with previous work}

The effect of feedback has been studied by numerous workers, both
analytically and numerically \citep[see, e.g., the review by][and
references therein]{VS11}. In particular, the pioneering numerical
simulations of \citet{BL80} included various cases of heating and
cooling functions for the medium, and used radiative transfer (on a $40
\times 40$ two-dimensional grid) to include the effects of
photoionization from OB stars in a 180-pc square region, making them a
direct precursor of this work. Even at their very limited resolution,
they foresaw several outcomes of this setup, such as the formation and
maintenance of a cloud population, that the clouds would be
gravitationally unstable had self-gravity been included, and that the
SFR would be self-consistent if 0.1--0.5 of the mass in the clouds were
to go into the formation of new massive stars, thus making a prediction
for the SFE. However, their required SFE for self-consistency was too
high, presumably because self-gravity was not included in their
simulations. This limitation also meant that the feedback stars had to
be placed randomly in the simulation.  Subsequent numerical works
focused mostly on the effect of supernova feedback on the structuring of
the ISM on kiloparsec scales \citep[e.g.,][] {Rosen+93, RB95,
deAvillez00, MacLow+05, JM06, Wood+10, Hill+12} but, in general,
self-gravity has not been included in these works, and the supernova
rate has been an input parameter for the simulations, rather than a
self-consistent output.

Another line of study has been the simulation of feedback by stellar
outflows at the clump (parsec) scale, aiming at either maintenance of
the turbulence within the clumps \citep[e.g.,][] {LN06, Carroll+09,
Cunningham+09}, or at the self-regulation of star formation
\citep[e.g.,][] {LN06, NL07, Wang+10}. The latter are closest in aim to
our present study, although not in scale, as they only consider
parsec-sized regions, and only outflow feedback, which corresponds to
the effect of low- and intermediate-mass stars, which does not seem to
be the dominant driver at the GMC scale \citep{Matzner02}.

Our results in this paper are most directly comparable to those of
\citet{Dale+12}, who performed a parameter-space study of the disruptive
effect of photoionizing radiation of molecular clouds of various masses.
Thus, the basic physical processes at play in their simulations are very
similar to those included in ours. Their main result is that, while
clouds of masses up to $\sim 10^5 ~\Msun$ can be readily destroyed by
the ionization feedback from their newly-formed stars, clouds with $M
\sim 10^6~\Msun$ cannot be destroyed, as their escape velocities are
larger than the sound speed in the photoionized gas. This regime is not
sampled by our simulations, in which the total dense gas mass is never
more massive than a few times $10^4~\Msun$.

The main differences between our setup and theirs are that they use a
polytropic equation of state covering only a temperature range
corresponding to molecular and cold atomic gas, and that they start with
a suite of initial clouds in various configurations, while we let the
clouds form self-consistently out of the warm ISM. Also, they include
the photoionized gas resulting from the stellar feedback, but their
simulations lack the warm (neutral and ionized) substrate in which the
clouds dwell, and out of which they form in our simulations. That is, in
their simulations there is no possibility of the warm environment
penetrating into the clouds, as proposed theoretically by \citet{HI06},
and suggested observationally by \citet{Krco+08}. Thus, our
self-consistently formed clouds may be more porous, and thus less bound,
than those of \citet{Dale+12}. Also, our probabilistic SF prescription
allows our SPs to be individual stars always and with a realistic
IMF. This means that our simulations can be employed for future cluster
dynamics studies. Instead, the sinks in the simulations by
\citet{Dale+12} have a mass range that goes from individual stars to
small clusters. Moreover, they only considered the effect of stars more
massive than $20~\Msun$, so they did not investigate the effect of stars
of different masses. On the other hand, their radiative transfer
algorithm is more realistic than ours, and they sample a larger
parameter space. Thus, in general, it can be said that the two studies
are highly complementary in nature, each one providing a different
perspective of the problem: specifically, they focused on the ability of
photoinizing radiation to destroy clouds of different masses, while we
have focused on the control of the SFE and the role of stars of different
masses and different feedback prescriptions. 

Most importantly, our simulations are relevant in the context of
studying the entire evolutionary cycle of molecular clouds, and showing
that star-forming GMCs as a whole can be in a global state of
gravitational contraction, which is initiated during their pre-molecular
stages, and yet comply with the low observed SFR and SFE in The Galaxy,
as a consequence of the stellar feedback, but not by maintaining them
hovering around an equilibrium state \citep{Krumholz06,
Goldbaum+11}, but rather by photoevaporating them while the collapse
motions continue.

\subsection{Interpretation and implications} \label{sec:interpretation}

\subsubsection{Evolution of the SFR and SFE} \label{sec:disc_SFR}

Our result that massive stars take several
megayears to form after the onset of SF, together with the fact that it
is the feedback from the massive stars that regulates the SFR, could be
naively taken to imply that the SFR (or the SFE) should be at its maximum at
the earliest stages of the clouds' evolution. However, this is not so
because {\it the clouds are evolving}. As observed in all simulations of
this process including self-gravity \citep[e.g.,][]{VS_etal07,
VS_etal10, VS_etal11}, and shown in the bottom-right panels of Figs.\
\ref{fig:Evolution_Central_Cloud},
\ref{fig:Evolution_LAF1_Single_Cloud1}, and
\ref{fig:Evolution_LAF1_Single_Cloud2}, the SFR in the clouds starts at
very low values, and increases over time. This can be understood as
follows: assuming that the accretion-induced turbulence in the clouds
\citep{KI02, Heitsch_etal05, VS_etal06} produces a certain density
probability density function (PDF), typically of lognormal shape for the
nearly isothermal dense gas \citep{VS94, PV98}, then one can assume that
the material responsible for the {\it instantaneous} SFR is that at
sufficiently high densities that its free-fall time is much shorter than
the cloud's dynamical timescale. This is at the basis of several recent
models for the SFR \citep[e.g.][see also Federrath \& Klessen 2012]
{KM05, PN11, HC11, ZVC12}. 

However, if the whole cloud is udergoing
global collapse, then its mean density is increasing, and its average
Jeans mass is decreasing, so that the fraction of its mass involved in
the instantaneous SF is also increasing, implying that the SFR increases
in time \citep{ZVC12}. In this context, once a sufficiently large dense
gas mass has been converted into stars, the IMF is expected to be
sufficiently sampled to produce massive stars which then begin to erode
the cloud and reduce the SFR again. This explains the fact that in
Figs.\ \ref{fig:Evolution_Central_Cloud},
\ref{fig:Evolution_LAF1_Single_Cloud1}, and
\ref{fig:Evolution_LAF1_Single_Cloud2}, the SFR first increases, reaches
a maximum, and finally begins to decrease again. In contrast, in
simulations with no feedback \citep[e.g.,][]{VS_etal07, VS_etal11}, the
SFR continues to increase until the dense gas mass is nearly exhausted.

The evolution of the SFE also deserves discussion. We note that the SFE
for the whole numerical box saturates at levels $\sim 10\%$ in both the
SAF1 and LAF1 runs (top right panels of Figs.\
\ref{fig:Evolution_SAF1_Box} and \ref{fig:Evolution_LAF1_Single_Box}).
These are the absolute efficiencies reached in the simulations. While
they may seem a bit high compared to standard estimates ($\sim 2\%$),
two factors should be considered. First, these are the {\it final}
efficiencies, which are largely observationally unconstrained, since it
is very difficult to know how much gas mass went into a SF episode after
no gas is left around a cluster. But it should be noticed that by the
time these runs form the most massive stars (a 20-$\Msun$ star at
$t=22.4$ Myr in SAF1 and a 30-$\Msun$ star at $t=23.5$ Myr in LAF1),
which could be considered to correspond to the typical observation of
the SFE in a GMC, the SFE in both runs is at the 1\% level. Second, in
any case, our simulations have neglected magnetic fields, which are
expected to reduce the SFE even if the clouds are magnetically
supercritical \citep[e.g.,][] {VS+05, VS_etal11, NL05}.

Another interesting issue about the SFE is that, for the Central Cloud
of run SAF1, it reaches a rather large value of $\sim 30\%$. This level
corresponds to that of cluster-forming clumps\citep[e.g.,][]{LL03}.
Instead, the SFE of Clouds 1 and 2 only reaches $\sim 10\%$ and $\sim
15\%$, respectively. This is all the more interesting because it can be
seen, from the top left panels of Figs.\
\ref{fig:Evolution_Central_Cloud} and
\ref{fig:Evolution_LAF1_Single_Cloud2}, that the maximum dense gas
masses of the Central Cloud and of Cloud 2 are very similar. The higher
SFE of the Central Cloud must then be attributed to the more strongly
focused character of the collapsing flow in run SAF1, which produces a
more compact cloud. Indeed, we have measured the evolution of the
gravitational potential for each cloud between the time of the onset of
SF and the time when the mass is dispersed, finding that the ratios of
the temporal minima of the bottom the potential wells of the Central
Cloud to that of Clouds 1 and 2 are both similar and about 1.5. This
indicates a higher degree of concentration of the Central Cloud compared
to the other two clouds, and suggests that the particular features of
the flow may have a significant influence on the SFE, besides the mass,
size, and velocity dispersion of the region.

\subsubsection{Evolution of the virial parameter}

A second important point to note is that our self-consistent {\it
evolutionary} simulations show that the virial parameter of the dense
gas takes values close to unity {\it before} stellar feedback is
dominant, while it takes much larger values once the feedback becomes
dominant. This is contrary to the common notion that stellar feedback
drives the turbulence in the clouds, maintaining them in approximate
equilibrium. Instead, our simulations suggest that the clouds take
values of $\alpha$ close to unity while they are dominated by
gravitational infall, and then take much larger values when they are in
the process of destruction by the feedback. 

This is actually consistent with the fact that GMCs tend to have masses
close to $\Mvir$ \citep[e.g.,][]{Heyer+09}, since by the time our clouds
take much larger values of this parameter, they do not appear as large
GMCs anymore, but rather as evacuated regions surrounded by cloud
shreds. This is seen, for example, in Fig.\ \ref{fig:images_cloud2}, in
the panels corresponding to times $t=25.6$, 27.0, and 27.6 Myr (top
right, middle left and middle right, respectively), noting that the
first very massive star ($30 \Msun$) appears at $t = 23.5$ Myr. This
suggests that the star-forming GMCs are in general still dominated by
the gravitationally collapsing motions. This result is also consistent
with the result by \citet{Dobbs+11} that clouds in Galactic-scale
simulations tend to have $\alpha$ distributions around unity when the
stellar feedback is artificially set to be very inefficient, while
clouds in simulations with larger feedback efficiencies tend to have
distributions of $\langle \alpha \rangle > 1 $.

At this point, it is important to note that \citet{Dobbs+11} interpret
the \citet{Heyer+09} data as meaning that most clouds are
gravitationally unbound, with $\alpha > 1$. However, the latter authors
themselves interpret their data as implying that $\alpha \sim 1$ on
average, in particular because their methods are likely to have
introduced an underestimation of the clouds' masses by factors of
2--3. Other clump surveys for which mass determinations independent of
the virial mass exist are consistent with the nearly-virialized (or,
alernatively, free-falling) state of the clouds \citep[see the
compilation by][] {BP+11}.

\subsection{Limitations} \label{sec:limitations}

Our simulations, although one step ahead of our previous effort from
Paper I, are still far from being all-inclusive. Most notably, we have
neglected supernova explosions and magnetic fields, and moreover, our
radiative transfer scheme is very rudimentary. We plan to improve on
these issues in future work. Here we can speculate what should be the
effect of these processes on our results.

First, as already mentioned in Sec.\ \ref{sec:disc_SFR}, our
simulations have neglected the effect of magnetic fields, supernova
explosions, and radiation pressure, all of which are expected to provide
additional regulation of the SFR and the SFE.


Second, concerning the radiative transfer (RT), our rudimentary PMRT
scheme does not account for the real column density between an ionizing
source and the test grid cell to be ionized, but only aims to represent
it by taking the geometric mean of the density at the source and at the
test cell. Thus, if a dense clump lies between these two cells, our
scheme will miss it, together with any shadowing effect it may
have. Thus, our scheme may tend to overestimate the photoionized
volume. We do not expect this effect to be dominant, since the volume
covered by shadows is not large, but comparisons should be performed
once a more thorough RT algorithm is implemented.

Finally, an important limitation is that we have considered only
relatively low-mass clouds. Sub-Galactic-scale numerical simulations
that have studied the feedback from massive stars on more massive clouds
have either not considered the evolutionary process leading to their
formation and self-consistent internal levels of turbulence
\citep[e.g.][]{Dale+12, Dale+13}, or else have neglected the
self-gravity of the gas \citep[e.g.,][]{KT12}. Therefore, the
investigation of this problem in the framework of the self-consistent
evolution of the clouds from their formation to their destruction
remains an open problem.

\section{Summary and conclusions} \label{sec:conclusions}

In this paper we have presented a numerical study of the entire
evolutionary cycle of molecular clouds, starting from their formation by
converging flows in the warm ISM, and concluding with their dispersion
by the photoinization feedback from the massive stars formed within
them. Our study extended the one presented in Paper I in two main areas.
First, we included a probabilistic scheme for star formation, which
serendipitously allowed us to produce a stellar population following a
realistic IMF, in turn allowing us to overcome a shortcoming of Paper I,
namely that all stellar particles (SPs) radiated with the same
intensity, roughly corresponding to that of a $\sim 10$--$\Msun$ star.
Second, we introduced our ``poor man's radiative transfer scheme'', PMRT,
which allowed us to produce mass-dependent Str\"omgren spheres, and thus
allowing to study the effect of stars of different masses in the
dispersal of their parent clouds.

We performed numerical simulations with initial conditions identical to
those used in Paper I, but varying the feedback schemes, in order to
quantify the difference between our old and new feedback schemes (PMRT
{\it versus} dumping all the energy in a single grid cell) and the
effect of including stelar populations of different mass ranges. The
simulation with small-amplitude initial velocity fluctuations, SAF1 (see
Table \ref{tab:run_params}), due to the larger coherence of the
converging motions in the warm gas, leads to the formation of a single,
more massive cloud at the center of the grid, which we called ``The
Central Cloud''. Instead, the simulations with large-amplitude initial
fluctuations, generically denoted LAF, produce various less-massive clouds
in the numerical box, away from the center of the simulation. From these
we selected two, which we labeled Clouds 1 and 2. 

We showed that cylindrical regions of length and diameter equal to 10 pc
were completely evacuated of dense gas on timescales $\sim 10$ Myr when
a full IMF was included in the calculations, and in fact, the total
dense gas mass in the numerical box is reduced by a factor $\sim 5$ in
the SAF1 simulation within $\sim 20$ Myr, and by a factor $\sim 10$ in
run LAF1 within $\sim 15$ Myr. Instead, when the most massive stars ($M
> 8~\Msun$) are not included in the simulations (runs LAFold and LAF8),
the total dense gas mass in the simulations is hardly affected, although
the SFE is reduced to levels $\sim 20\%$. When stars up to $20 ~\Msun$
are included (run LAF20), the total dense gas mass in the simulation is
reduced at a level comparable to that of run LAF1, but on a timescale
almost twice as long. Thus, our results strongly suggest that the
destruction of the clouds is accomplished by stars with masses $M \simgreat
20~\Msun$.

Our simulations also show that star formation events can be
completely terminated, and the dense gas completely dispersed, on scales
$\simless 10$ pc by the photoionizing effect of the newly formed stars
in those regions, while at larger scales the dense gas contents is
decreased but not completely destroyed, and the SFR is analogously
decreased but not terminated. This suggests that the stellar
photoionizing feedback can locally disrupt the clouds and terminate SF,
but new SF events can occur later at new locations in the clouds.

We have also investigated the evolution of the virial parameter of
the clouds, finding that it approaches unity {\it before} the stellar
feedback begins to dominate the dynamics; that is, while the clouds are
dominated by the infalling motions that drive their growth. Later, when
the feedback becomes dominant, the clouds are eroded away by the
ionisation heating, and the virial parameter increases, both because
the heating induces expanding motions in the gas, and because the dense
gas mass decreases as it is evaporated away. This suggests that the
clouds become unbound {\it as a consequence} of the stellar feedback,
rather than the unboundness being the cause of a low SFE, as has been
recently suggested by \citet{Dobbs+11}.

Finally, a collateral result is that chains of isolated dense blobs,
resembling those in the vicinity of the famous dark globule B68, are
formed as the filaments feeding the cluster-forming clumps are eroded by
the ionisation heating from the massive stars. This occurs because the
filaments are themselves clumpy, and these clumps survive the ionisation
heating for longer times than the rest of the filaments.

In conclusion, our simulations show that the scenario in which large, dense
and cold clouds begin to collapse even before they are mostly molecular,
and continue doing so through their star-forming stages is 
consistent with the observed values of the SFE and with the morphology
of the clouds.

\section*{Acknowledgments}
We are grateful to A. Kravtsov for providing us with the numerical code,
and to N. Gnedin for providing us with the useful graphics package IFRIT. We
acknowledge the anonymous referee, whose insightful comments and questions
improved various aspects of this paper. This work has received partial financial
support from CONACYT grant 102488 to E.V.-S. and UNAM-DGAPA PAPIIT grant
IN106511 to GCG.

\label{lastpage}

\end{document}